\documentclass[conference]{IEEEtran}
\IEEEoverridecommandlockouts

\usepackage{cite}
\usepackage{amsmath,amssymb,amsfonts}
\def\BibTeX{{\rm B\kern-.05em{\sc i\kern-.025em b}\kern-.08em
    T\kern-.1667em\lower.7ex\hbox{E}\kern-.125emX}}

\usepackage{url}
\usepackage{algorithm,algpseudocode}
\usepackage{siunitx}
\usepackage{multirow}
\usepackage{tikz}
\usepackage{pgfplots}
\usepackage{csquotes}
\pgfplotsset{compat=1.18}
\usetikzlibrary{calc}

\usepackage[hidelinks]{hyperref}

\newcommand{\circumeq}{\mathrel{\hat{=}}}
\newcommand{\code}[1]{\texttt{#1}}

\definecolor{clrTime}{RGB}{212,175,55}
\definecolor{clrDevice}{RGB}{135,133,0}
\definecolor{clrZL}{RGB}{0,0,0}

\definecolor{clrHEFT}{RGB}{77,175,74}
\definecolor{clrPEFT}{RGB}{166,86,40}

\definecolor{clrSA}{RGB}{0,0,255}
\definecolor{clrNSGA}{RGB}{0,0,0}

\definecolor{clrSN}{RGB}{228,26,28}
\definecolor{clrSNFF}{RGB}{255,127,0}
\definecolor{clrSP}{RGB}{55,126,184}
\definecolor{clrSPFF}{RGB}{152,78,163}

\begin{document}

\title{Static task mapping for heterogeneous systems based on series-parallel decompositions}


\author{\IEEEauthorblockN{1\textsuperscript{st} Martin Wilhelm}
\IEEEauthorblockA{\textit{Otto-von-Guericke University}\\
Magdeburg, Germany \\
martin.wilhelm@ovgu.de}
\and
\IEEEauthorblockN{2\textsuperscript{nd} Thilo Pionteck}
\IEEEauthorblockA{\textit{Otto-von-Guericke University}\\
Magdeburg, Germany \\
thilo.pionteck@ovgu.de}
}

\maketitle

\begin{abstract}
Modern heterogeneous systems consist of many different processing units, such as CPUs, GPUs, FPGAs and AI units. A central problem in the design of applications in this environment is to find a beneficial mapping of tasks to processing units. While there are various approaches to task mapping, few can deal with high heterogeneity or applications with a high number of tasks and many dependencies. In addition, streaming aspects of FPGAs are generally not considered. We present a new general task mapping principle based on graph decompositions and model-based evaluation that can find beneficial mappings regardless of the complexity of the scenario. We apply this principle to create a high-quality and reasonably efficient task mapping algorithm using series-parallel decompositions. For this, we present a new algorithm to compute a forest of series-parallel decomposition trees for general DAGs. We compare our decomposition-based mapping algorithm with three mixed-integer linear programs, one genetic algorithm and two variations of the Heterogeneous Earliest Finish Time (HEFT) algorithm. We show that our approach can generate mappings that lead to substantially higher makespan improvements than the HEFT variations in complex environments while being orders of magnitude faster than a mapper based on genetic algorithms or integer linear programs.
\end{abstract}


\section{Introduction}


Modern application systems are becoming increasingly heterogeneous as they consist of multiple components that can be accelerated using specialized processors, such as GPUs, programmable logic or AI units. If multiple components are eligible for acceleration, the question arises how to map these components (tasks) to the available accelerators. This process is generally referred to as \emph{task mapping}, where applications are represented as a directed acyclic \emph{task graph} indicating the data dependencies between various tasks. Finding an optimal task mapping is non-trivial. Due to data transfer costs from and to an accelerator, the benefit of the mapping of one task highly depends on the mapping of its neighbors. Furthermore, the mapping of multiple independent tasks to the same accelerator may lead to contention, leaving other parts of the system idle. 

While a significant body of research exists for task mapping in heterogeneous systems, most algorithms are designed with a focus on systems with low heterogeneity, such as CPU clusters or CPU-GPU systems, and task graphs with few nodes or few dependencies between these nodes. As a consequence, many existing static task mapping algorithms do not scale well to complex graphs, which consist of both a high number of tasks and a high number of dependencies between them. Furthermore, dataflow streaming on an FPGA along different tasks is generally not considered. Algorithms that are still able to find significant improvements in such scenarios are usually based on integer linear programming or metaheuristics and, consequently, have a high computational complexity.


In this work, we present a decomposition-based task mapping principle, which leads to scalable and efficient task mapping heuristics in arbitrarily complex heterogeneous systems. This principle is based on the fundamental observation that complex task graphs may be decomposed into various subgraphs that can be beneficially mapped onto the same accelerator, thereby enabling synergies between these nodes such as reduced data transfer times or potential dataflow streaming. A promising candidate for such decompositions are series-parallel graphs, which can be naturally decomposed into largely independent subgraphs. We use this characteristic to substantially improve the quality of task mappings for a large range of potential applications. Our algorithmic principle is designed as a model-based approach, that is, we build upon a well-defined system model, which allows us to make predictions about the expected makespan. Through repeated re-evaluation, we exploit the fact that in a model-based environment we can efficiently evaluate and compare any given mapping at design time.

In the following section, we give a short introduction to static task mapping and series-parallel graphs. In Section~\ref{sec:decompositionmapping}, we introduce the decomposition mapping principle and provide two task mapping algorithms based on this principle. Furthermore, we propose a heuristical optimization of the general principle, which leads to a significantly reduced execution time. In Section~\ref{sec:evaluation}, we compare our approach to the two well-known list scheduling algorithms HEFT and PEFT, a genetic algorithm and three different integer linear programs. For this, we use randomly generated task graphs as well as task graphs that are extracted from realistic workflows.

\section{Background}

\subsection{Static task mapping}


Task mapping refers to the process of assigning (sub)tasks of an application to processing units of a heterogeneous platform. 
Task mapping approaches can generally be divided into \emph{static} and \emph{dynamic} task mapping~\cite{mittal2015}. Static approaches generate a full task mapping for the system before the execution based on profiling data or performance models, whereas dynamic approaches can take runtime information into account. Although more flexible, dynamic task mapping requires predefined implementations of all tasks on all available processing units, which makes it most suited for systems with a small number of recurring jobs and few different processing units. In this work, we put a focus on complex environments and therefore on static task mapping.

Zhou and Liu use a mixed-integer linear program to find an optimal static mapping and compare its results with two dynamic mapping heuristics~\cite{zhou2014}. Similarly, there is various recent work creating integer linear programs for the mapping problem~\cite{emeretlis2022,mohammadi2023,wilhelm2023}. 
A well-known heuristic from the field of task scheduling is the Heterogeneous Earliest Finish Time (HEFT) algorithm, which creates a task mapping together with a schedule~\cite{topcuoglu2002}. HEFT performs very well in a CPU-GPU environment, but has a mostly local view on the task graph in its scheduling phase. There are many list scheduling algorithms similar to HEFT with modifications that aim to mitigate this downside~\cite{bittercourt2010,arabnejad2014,sirisha2023_2}. Maurya et al.\ present a comparison of various heterogeneous list scheduling algorithms, showing that they generally lead to similar performance with small differences depending on the application scenario~\cite{maurya2018}. They found that the Predict Earliest Finish Time (PEFT) algorithm~\cite{arabnejad2014} performs slightly better than the other algorithms for bigger graphs and a more heterogeneous environment.

In the context of multiprocessor system-on-chips (MPSoCs), a large body of research on static task mapping exists, which is dominated by metaheuristic approaches~\cite{singh2013, gupta2021}. 
Especially genetic algorithms are consistently found to be very effective in finding profitable mappings~\cite{braun2001}. 
Most genetic algorithms used in the context of task mapping are based on NSGA-II~\cite{deb2000} or SPEA2~\cite{zitzler2001}, which are designed to find pareto-optimal solutions in multi-objective optimization problems. 
In this work, we focus on a single objective makespan optimization. However, the basic algorithmic ideas presented in this work can easily be transferred to multi-objective optimization.

\subsection{Model-based evaluation}

As of today, there is no unified method on comparing task mapping algorithms. Experiments are usually run with custom implementations on specialized platforms or custom simulators, which makes it hard to transfer results. One approach to make results more comparable is to use an open-source simulation framework~\cite{gries2004,ricogallego2019}. 
Simulation frameworks, however, have the tendency to become very complex in order to accurately reflect the system behavior. This leads to many design decisions in the simulators that are coupled to the expected hardware and make it hard to compare generalized algorithms for task mapping using these frameworks~\cite{goens2016}.


A recent approach on the design and evaluation of heterogeneous task mapping algorithms is a fully model-based environment~\cite{wilhelm2022}. Here, the authors suggest using a fast and fully model-based cost function to evaluate the overall cost of the execution. For a comparison of mapping algorithms, it is not necessary to be able to determine the exact execution time of a system as long as the model behaves the same as the actual system. This follows the design philosophy of computational complexity theory, where it is more important to have the ability to analyze the behavior of algorithms than to predict their exact execution times. Nevertheless, in~\cite{wilhelm2023}, it was shown that with the presented model the actual execution time can be closely predicted for basic tasks with error rates of less than~\SI{10}{\percent}. Adapting this approach bears many advantages. First, it enables the analysis of a high number of arbitrarily complex use cases during the evaluation. Second, we can make use of the fact that we have access to a fast cost function during algorithm design. In particular, this enables the evaluation of intermediate results during the construction of a task mapping. In this work, we exploit this fact to efficiently find profitable task mappings.

\subsection{Series-parallel graphs}

\begin{figure}[ht]
\centering
\includegraphics[width=0.9\linewidth]{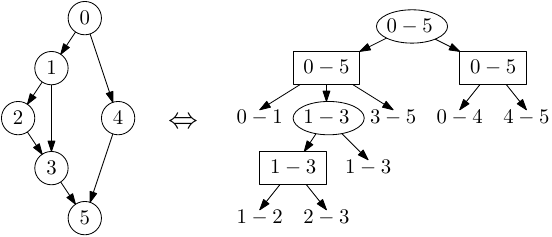}
\caption{A directed series-parallel graph and its decomposition tree. A round node indicates a parallel operation, whereas a rectangular node indicates a series operation. The leaves represent an edge in the original graph.}
\label{fig:sp_decomposition}
\end{figure}

Series-parallel graphs are a well-known graph class, originally described for the analysis of electrical networks~\cite{duffin1965}. A series-parallel graph is recursively defined as either two nodes connected by a single edge (a $K_2$) or a graph that results from another series-parallel graph via a \emph{series} or a \emph{parallel} operation. Here, a series operation refers to the insertion of a node on an edge (which splits the edge into two edges), whereas a parallel operation refers to the duplication of one edge. Analogously, directed acyclic series-parallel-graphs can be defined, starting from a directed edge. By construction, series-parallel graphs are always planar and, hence, have a linear complexity. Furthermore, they allow for the computation of a series-parallel decomposition tree, which divides the graph into largely independent subgraphs~\cite{eppstein1992}. An exemplary decomposition is shown in Fig.~\ref{fig:sp_decomposition}. Here, the root node represents the parallel operation between start and end node, which divides the graph into the left and the right subgraph. Its child nodes then represent the series operations that are further dividing these subgraphs. The leaves of the decomposition tree are equivalent to edges in the original graph.

Intuitively, series-parallel graphs define structures that can be arbitrarily decomposed into substructures with well-defined input and output. If a series-parallel graph is interpreted as a task graph, this means that a graph is series-parallel if and only if subfunctionalities (i.e.\ subgraphs) do not access intermediate values of other subfunctionalities. This property equates to the principle of modular programming, which is a fundamental principle in modern software engineering~\cite{cai2023}. Consequently, many real-world applications can be expected to yield (almost) series-parallel task graph representations.

\section{Decomposition-based mapping}
\label{sec:decompositionmapping}

In this section, we present a new heterogeneous task mapping principle based on graph decompositions and model-based evaluation. We first describe the general principle and afterwards present two heuristics following this approach. Finally, in Section~\ref{ssc:optimizations}, we discuss an optimization to the general principle, allowing for a tradeoff between execution time and quality of the result. In order to avoid confusion, we will refer to the time needed for the execution of the mapping algorithm as the \emph{execution time} and to the time needed to execute the accelerated application as the \emph{makespan}.

\subsection{General principle}
\label{ssc:decomp_principle}


The basic idea for decomposition-based mapping is to create a task mapping in a greedy way following four steps: 
\begin{enumerate}
\item Create a default mapping, where every task is mapped to a default device (usually a CPU).
\item Search for the subgraph whose acceleration on one of the available processing units has the highest impact on the overall makespan.
\item Map this subgraph to the respective processing unit. 
\item Repeat steps two and three until the makespan cannot be further reduced.
\end{enumerate}

While fairly straightforward, the realization of these steps bears two significant challenges. First, determining the impact of a change in the mapping on the overall makespan is hard without letting the actual system run on the target platform, which is clearly too expensive to be practical. Second, for a task graph of size $n$ there are $2^n$ possible subgraphs. Testing all subgraph replacements therefore takes $m2^n$ steps, where $m$ is the number of available processing units. Hence, a smart choice of the set of subgraphs is essential to reduce the execution time while maintaining a high result quality.

A natural way to compute the impact of a mapping change on the makespan is to use a constant-speed metric such as the raw acceleration factor combined with the cost for additional data transfers. This factor could also be weighted by the number of descendants or the projected length of the critical path starting from the replaced subgraph. While computing these metrics is generally very fast, they can only provide guesses based on a local view on the system. In consequence, decisions based on these metrics will often lead to a net increase in the actual makespan and, depending on their robustness, the algorithm may not terminate if there is a cycle of changes where each change predicts an improvement. Our solution to this problem lies in a full-scale model-based evaluation, which can be computed in linear time with respect to the number of edges~\cite{wilhelm2023}. That is, we fully re-evaluate the system for each subgraph replacement. With this, we are able to take a fully global view on the impact of a change and guarantee that a change leads to an improvement. Since the model-based evaluation is fully deterministic, we can furthermore guarantee that the algorithm terminates.

For finding a beneficial set of subgraphs, we analyze two different strategies in Sections~\ref{ssc:sndmapping} and~\ref{ssc:spdmapping}. A large number of subgraphs strongly increases the execution time of the algorithm, whereas a small number of subgraphs increases the chance that the algorithm may be stuck in a local minimum. We aim to find subgraph sets of complexity $O(n)$, which leads to a total execution time of $O(n^2m)$ for step~2 of the algorithm if the graph has $O(n)$ edges. Note that step~2 of the algorithm could still theoretically be executed $\Omega(m^n)$ times. However, in practice the number of iterations will almost always be $O(n)$. In order to avoid degenerate situations, an iteration cap of $n$ iterations may be introduced in a production environment.

\subsection{Single node decomposition}
\label{ssc:sndmapping}


The most basic approach to define a set of subgraphs with linear complexity is to choose the set of all nodes itself, i.e., all subgraphs consisting of a single node. This choice allows us in principle to reach every possible mapping with a minimal set of subgraphs. It can be expected that this approach works well in small to medium size task graphs with computation-heavy tasks, since there the mapping of a single node has a relatively high impact on the total makespan. With more complex task graphs and data-intensive applications, however, there is a high chance that this approach gets stuck in a local minimum where accelerating a single task does not outweigh the additional communication cost. This is especially true if FPGAs are involved, which can stream data along several connected tasks if all of them are mapped to the FPGA.

\subsection{Series-parallel decomposition}
\label{ssc:spdmapping}

A more sophisticated approach is to choose a subgraph set based on the series-parallel decomposition of a graph. The main idea behind this approach is to combine tasks into a subgraph if mapping all of them onto the same processing unit reduces data transfer overhead. The inner nodes of a series-parallel decomposition tree, as shown in Fig.~\ref{fig:sp_decomposition}, each represent a subgraph that has only one input and one output connection to the rest of the task graph. Building on this observation, we construct the set of subgraphs $S$ for the graph $G$ as follows:
\begin{enumerate}
\item Add all subgraphs consisting of a single node to $S$.
\item Compute a series-parallel decomposition tree $T$ of $G$.
\item For each series operation in $T$, add the subgraph of all nodes of this operation except start \& end node to $S$.
\item For each parallel operation in $T$, add the subgraph of all nodes of this operation including start \& end node to $S$. 
\end{enumerate}
The reasoning behind this choice of subgraphs is that in a series operation the start and end node may have other outgoing or incoming edges, respectively. Including these nodes in the subgraph would therefore possibly influence the communication cost of a sibling of the series operation. For a parallel operation, all children or parents, respectively, of its start and end node are included in the associated decomposition subtree, except for a few edge cases that may be avoided during the specification of the task graph. Therefore start and end node themselves act as single input/output source of the subgraph.
For the task graph shown in Fig.~\ref{fig:sp_decomposition}, this would result in the following set of subgraphs:
\[
S = \left\lbrace \lbrace 0\rbrace, \lbrace 1\rbrace, \lbrace 2\rbrace, \lbrace 3\rbrace, \lbrace 4\rbrace, \lbrace 5\rbrace, \lbrace 1,2,3\rbrace, \lbrace 0,1,2,3,4,5\rbrace \right\rbrace
\]
Note, that in this case the series operations yield either single-node subgraphs ($\lbrace 2\rbrace, \lbrace 4\rbrace$) or a subgraph that is identical to a subgraph defined by a parallel operation ($\lbrace 1,2,3\rbrace$).

Naturally, this approach only works for series-parallel graphs. While we expect that real-world application task graphs are often almost series-parallel, a single conflicting edge causes the above algorithm to fail. To circumvent this, we partition general DAGs into series-parallel subgraphs and create a forest of decomposition trees whose operations are then added to the subgraph set as before. 
Finding the maximal series-parallel subgraph of a general graph is NP-hard~\cite{calinescu2012}. To the best of our knowledge, there is currently no widely recognized algorithm to partition a DAG into series-parallel subgraphs. Hence, we provide an original algorithm.

\begin{algorithm}
\begin{algorithmic}[1]
\footnotesize
\Function{grow\_decomposition\_forest}{Start node $s$, Forest $\mathbb{F}$}
\State $T\gets\Call{grow\_series}{[\varepsilon, s],\mathbb{F}$}
\State{$\mathbb{F}=\mathbb{F}\cup \lbrace T\rbrace$}
\EndFunction\\

\Function{grow\_series}{Tree $T\circumeq [u,v]$, Forest $\mathbb{F}$}
\While {$v\neq\varepsilon$ \textbf{and} $\Call{indegree}{v} \leq \Call{outsize}{T}$}		\label{line:check_additional_edges}
	\If{$\Call{outdegree}{v}=1$}
		\State $w\gets\Call{successor}{v}$
		\State $T=\Call{series}{T,[v,w]}$	\label{line:single_succ}
	\Else
		\State $T_p=\Call{grow\_parallel}{v,\mathbb{F}}$	\label{line:recursive_grow}
		\State $T=\Call{series}{T,T_p}$		\label{line:extend_parallel}
	\EndIf		
\EndWhile
\State\Return $T$
\EndFunction\\

\Function{grow\_parallel}{Node $v$, Forest $\mathbb{F}$}
\State $\mathbb{W}\gets$ Wavefront (empty forest)
\ForAll{$(v,w)\in\Call{outedges}{v}$}
\State $\mathbb{W} = \mathbb{W}\cup\lbrace [v,w]\rbrace$
\EndFor
\Loop
	\Repeat
		\While{$\exists\mathbb{W}'\subseteq\mathbb{W}: |\mathbb{W}'|\geq 2, \forall T'\in\mathbb{W}': T'\circumeq [u_1,u_2]$}\label{line:combine_parallel_begin}
			\State $\mathbb{W} = \mathbb{W}\setminus\mathbb{W}'\cup\lbrace\Call{parallel}{\mathbb{W}'}\rbrace$
		\EndWhile		\label{line:combine_parallel_end}
		\If{$\exists T: \mathbb{W} = \lbrace T\rbrace$} \label{line:parallel_return_begin}
			\State\Return $T$				\label{line:parallel_return_end}
		\Else
			\ForAll{$T\in\mathbb{W}$}
				\State $T \gets \Call{grow\_series}{T, \mathbb{F}$}				\label{line:grow_active_subtree}
			\EndFor 
		\EndIf
	\Until no change in the wavefront $\mathbb{W}$ occurred\\
	
	\State Choose any $T_c\in\mathbb{W}$ whereby $T_c\circumeq[u_1,u_2]$	\label{line:cuttree_begin}
	\State $\mathbb{F} = \mathbb{F}\cup\lbrace T_c\rbrace, \mathbb{W} = \mathbb{W}\setminus\lbrace T_c\rbrace$
	\State $\Call{indegree}{u_2} \gets \Call{indegree}{u_2} - \Call{outsize}{T_c}$							\label{line:cuttree_end}
\EndLoop
\EndFunction
\end{algorithmic}
\caption{Creating a forest of series-parallel decomposition trees for general DAGs with distinct start and end nodes.}
\label{alg:sppartition}
\end{algorithm}

In Alg.~\ref{alg:sppartition}, we successively grow decomposition trees with a series or parallel operation in their root starting from a single start node of the DAG and ending in a single end node. We can assume that the DAG has a single start and end node since otherwise we may just insert new start and end nodes with an edge to or from all existing start or end nodes, respectively. Each series-parallel decomposition tree $T$ represents a subgraph with distinct start and end nodes $u,v$ and can therefore be treated equivalently to an edge $(u,v)$. We use the notation $[u,v]$ to denote a decomposition tree consisting of a single edge $(u,v)$. We then use the notation $T\circumeq[u,v]$ to reflect that the subgraphs represented by $T$ and $[u,v]$ have the same start and end nodes. The main method \code{grow\_decomposition\_forest} starts the algorithm by growing a series operation starting from a virtual incoming edge $(\varepsilon, s)$ to the start node $s$. The resulting decomposition tree represents the core decomposition tree, reaching from start to end edge, from which other branches may be cut off during its creation if the original graph is not series-parallel. In \code{grow\_series}, we aim to extend a series operation either by a single edge if the current end node has only one successor (line~\ref{line:single_succ}) or by a recursively defined parallel operation (line~\ref{line:recursive_grow}-\ref{line:extend_parallel}). This extension is only possible if all incoming edges of the current end node are part of the subgraph represented by $T$ (line~\ref{line:check_additional_edges}), since otherwise the current series operation is part of a larger parallel operation. To check this, we manage the \code{outsize} of each decomposition tree $T\circumeq[u,v]$, which is the number of edges in $T$ with endpoint $v$. Aside from that, growing the operation stops if the end of the DAG is reached, which is indicated by a virtual outgoing edge $(t,\varepsilon)$.

\begin{figure}[ht]
\centering
\includegraphics[width=\linewidth]{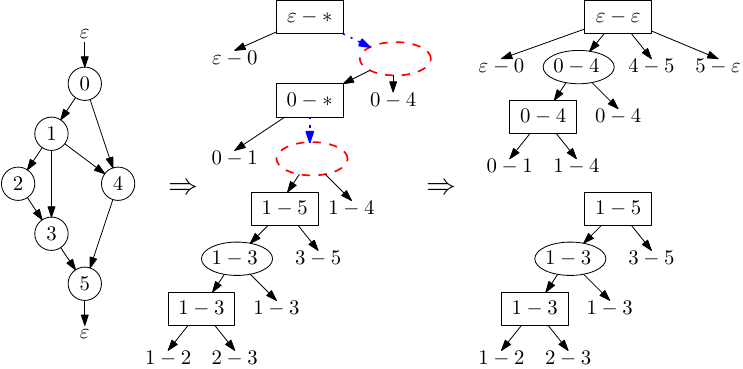}
\caption{An illustration of the cutting step of Alg.~\ref{alg:sppartition} and the resulting decomposition forest for a non-series-parallel graph. Blue dotted edges indicate a currently active \code{grow\_series} call whereas red dashed ellipses indicate an active \code{grow\_parallel} call. Stars in a series operation indicate that the final end node has not yet been found. The decomposition forest on the right side results from cutting the subgraph $1-5$.}
\label{fig:cuttingstep}
\end{figure}

In order to grow a subtree with a parallel operation, we maintain a wavefront $\mathbb{W}$, which contains all active subtrees. We then repeatedly grow all active subtrees (line~\ref{line:grow_active_subtree}) until two or more active subtrees have the same end node and therefore can be combined in a parallel operation (line~\ref{line:combine_parallel_begin}-\ref{line:combine_parallel_end}). If only one active subtree is left, the parallel operation is complete (line~\ref{line:parallel_return_begin}-\ref{line:parallel_return_end}). If the input graph is series-parallel, this will always happen after all possible changes to the wavefront have been made. Hence, if after one iteration of the loop there is no change in the wavefront, the graph is not series-parallel and one of the active decomposition trees represents a conflicting edge. In this case, we randomly choose one of the active decomposition trees and cut this tree $T_c$ from the DAG (line~\ref{line:cuttree_begin}-\ref{line:cuttree_end}). For this, it is sufficient to remove $T_c$ from the wavefront and to reduce the number of expected inputs at its end node. 
Fig.~\ref{fig:cuttingstep} shows a non-series-parallel graph, the intermediate state right before a subgraph has to be cut and a potential resulting forest. As indicated by blue dotted lines and red dashed ellipses, there are two calls to \code{grow\_series} and two calls to \code{grow\_parallel} active. The bottommost call to \code{grow\_parallel} happens at node~$1$. Here, both active branches do not lead to the same end node, but also cannot be further grown, since the branch $1-5$ is blocked by the edge $4-5$ and the branch $1-4$ is blocked by the edge $0-4$. With this, we know that the graph is not series-parallel and one of the two branches must be cut. The forest on the right depicts the situation where the branch $1-5$ has been cut. If the choice would fall to the branch $1-4$, the resulting forest would consist of the decomposition tree shown in Fig.~\ref{fig:sp_decomposition} and a single edge, which is arguably a better decomposition in the context of task mapping. A well-designed heuristic might exploit this observation in order to improve the resulting mapping algorithm.
With a careful implementation, the presented decomposition algorithm runs in linear time with respect to the number of edges of the original DAG.

\subsection{$\gamma$-Threshold and FirstFit mapping}
\label{ssc:optimizations}


The size of the decomposition affects the total execution time of the decomposition mapping algorithm multiplicatively as each possible replacement is evaluated in each iteration (cf.\ Sec.~\ref{ssc:decomp_principle}). However, many possible mapping operations are far from being useful, which makes it superfluous to evaluate mappings resulting from these operations in every iteration. Although generally, the exact same mapping will seldom be evaluated twice, we can make use of the fact that similar mappings have been evaluated in previous iterations of the algorithm. In the \emph{$\gamma$-threshold} variant of decomposition mapping, we assign an expected makespan improvement to each mapping operation after the first iteration of the algorithm and sort the mapping operations by this value using a priority queue. If we then find a mapping that leads to an improvement, we only look ahead for those mappings where the expected improvement is higher than the current improvement divided by~$\gamma$. If mappings are recomputed during this process, their respective expected improvements are updated in the priority queue. As before, the algorithm terminates if no improvement could be found. Therefore, in the last iteration, we recompute every possible mapping to make sure that initially bad mapping operators are at least recomputed once to evaluate their impact on the final configuration.

A special case of the $\gamma$-threshold decomposition mapping is the situation where $\gamma=1$. Here, generally, the first improvement that is found is applied to the graph, except if the improvement is significantly smaller than the previously expected improvement. We call this approach \emph{FirstFit decomposition mapping}. While less forgiving than a variant with $\gamma > 1$, this approach effectively minimizes the amount of required recomputations.

\section{Evaluation}
\label{sec:evaluation}

In this section, we aim to give an experimentally supported intuition on the advantages and disadvantages of the decomposition-based approach. For this, we compare our approach with various other task mapping algorithms using model-based evaluation. For the evaluation, we use both randomly generated graphs and real-world graphs.

\subsection{Evaluation system}


We compare different mapping strategies using model-based evaluation. We consider a platform consisting of one AMD Epyc 7351P CPU, one AMD Radeon RX Vega 56 GPU and one Xilinx XCZ7045 FPGA, which are characterized according to the platform model presented in~\cite{wilhelm2023}. We subsequently compare our approach with three different mixed-integer linear programs (MILPs), a genetic algorithm and two heterogenenous scheduling heuristics as listed below.\\

\textbf{ZhouLiu}  The MILP presented by Zhou and Liu~\cite{zhou2014} represents one of the first and most detailed MILPs for a CPU-GPU environment, which creates a total order of tasks on each processing unit by assigning execution slots to each task. It can be expected to produce very good results at high computation cost.\\

\textbf{WGDP\_Dev / WGDP\_Time} The device-based MILP by Wilhelm et al.~\cite{wilhelm2023} represents a family of similar MILPs, which aim to balance the workload on the available processing units without considering dependencies. While very fast compared to other MILP-based approaches, it may lead to worse results in complex task graphs with many dependencies. The time-based MILP proposed by the same authors aims to order tasks by determining start and finish times for all tasks. As it is the only MILP that takes data streaming into account, it can be expected to produce better results than other MILPs in a system where FPGAs are present.\\

\textbf{HEFT} The Heterogeneous Earliest Finish Time (HEFT) algorithm introduced by Topcuoglu et al.~\cite{topcuoglu2002} is the main representative of a family of priority-based list scheduling algorithms for heterogeneous environments. It computes an EFT for all tasks based on the average computation time, the average data transfer time and the maximum EFT of its successors. Afterwards, tasks are scheduled insertion-based in order of their EFTs.\\

\textbf{PEFT} The Predict Earliest Finish Time (PEFT) algorithm by Arabnejad and Barbosa~\cite{arabnejad2014} is similar to HEFT, but uses a so-called optimistic cost table (OCT), which assigns EFTs to task-device pairs, in order to get a more accurate EFT. With this, it is able to compute EFTs using the minimum instead of the average EFT of its successors. It is shown to be one of the best-performing HEFT variants for complex systems~\cite{maurya2018}.\\

\textbf{NSGAII} Genetic algorithms have proven to be a very effective approach for task mapping in heterogeneous systems, especially in the context of MPSoCs. We use a single objective variant of the NSGA-II algorithm~\cite{deb2000}. We employ a single-point crossover with a crossover rate of \SI{90}{\percent} based on a topologically sorted genome with one gene for each task. We use a mutation rate of $\frac{1}{n}$, where $n$ is the length of the genome, and a population size of $100$ individuals. We use a repair function after crossover to ensure that the produced mappings are feasible. If not stated otherwise, we execute the algorithm for \num{500} generations.\\

From the considerations made in Section~\ref{sec:decompositionmapping}, we derive four different decomposition-based mapping algorithms:\\

\textbf{SingleNode / SNFirstFit} The single node decomposition mapping algorithm as introduced in Section~\ref{ssc:sndmapping} without and with the FirstFit heuristic described in Section~\ref{ssc:optimizations}.\\

\textbf{SeriesParallel / SPFirstFit} The series-parallel decompo-sition mapping algorithm as introduced in Section~\ref{ssc:spdmapping} without and with the FirstFit heuristic.\\

For the evaluation, we determine the average over $30$ randomly generated or randomly augmented graphs of the respective size for each data point if not stated otherwise. We apply the cost function introduced by Wilhelm et al.~\cite{wilhelm2023} with FPGA streaming support. For each graph, we determine the makespan of a mapping as the minimum among all makespans that are computed using a breadth-first schedule and \num{100} randomly generated schedules. Based on the makespans of the different graphs, we generally compute the \emph{average positive relative improvement} of the makespan, i.e., the average relative improvement over a pure CPU mapping, whereas we count deteriorations as zero improvements. The proposed decomposition-based mapping algorithms are by design never worse than a pure CPU mapping. As one can always default to a pure CPU mapping, truncating negative improvements for the reference algorithms results in a fairer comparison. We refer to this value shortly as the \emph{relative improvement}.

Our tests are executed on an AMD Epyc 7351P CPU with \SI{2}{\tera\byte} RAM. The evaluation environment is written in C++20 and compiled using g++~9.4.0. The mixed-integer linear programs are solved using the Gurobi Optimizer v9.1.2~\cite{gurobi}.

\subsection{Random series-parallel graphs}
\label{ssc:seriesparallel}

For a basic evaluation, we use a test dataset consisting of random series-parallel graphs. The graphs are generated by starting with a single directed edge and randomly executing series or parallel operations in a ratio of 1:2 until the desired number of task nodes is created. Finally, redundant edges are removed from the resulting DAG. The generation is biased towards parallel operations since many of them insert edges that are deleted in the end.
The graphs are augmented with random complexity, parallelizability and streamability parameters. The complexity determines the number of operations per data point for a task, whereas parallelizability and streamability determine how well a task can be accelerated by multicore CPUs, GPUs and/or FPGAs  (see~\cite{wilhelm2023}). Both complexity and streamability are generated using a lognormal distribution with $\mu=2$ and $\sigma=0.5$. With this, \SI{90}{\percent} of the values are in the range from $3$ to $17$ with a median of about $7.4$. For parallelizability, we take into account that, by Amdahl's law, the potential performance gain through parallelization rapidly declines if tasks are not perfectly parallelizable. Hence, we assign a perfect parallelizability to a task with \SI{50}{\percent} probability and assign a uniform random parallelizability in the range of \SIrange{0}{100}{\percent} to the rest. For mapping on an FPGA, we assign an area limitation proportionally to the task's complexity. For the data transfer, we assume a constant data flow of \SI{100}{\mega\byte} between tasks.

\begin{figure}[ht]
\centering
\begin{tikzpicture}[trim axis left, trim axis right]
\begin{axis}[height=0.23\textwidth,width=0.95\linewidth,legend to name=milpimprlegend,legend cell align={left},legend columns = 3,legend style={font=\scriptsize},label style={font=\footnotesize},ylabel={Relative improvement},every axis plot/.append style={thick,mark options={scale=0.5}, mark repeat=17},xticklabel=\empty] %

\addlegendentry{WGDP\_Time}
\addlegendentry{WGDP\_Device}
\addlegendentry{ZhouLiu}
\addlegendentry{SingleNode}
\addlegendentry{SeriesParallel}

\addplot[clrTime,mark=*, mark phase = 4] coordinates{(5,0.182476) (6,0.0930103) (7,0.137491) (8,0.101536) (9,0.119524) (10,0.123717) (11,0.160047) (12,0.139519) (13,0.171414) (14,0.204155) (15,0.202189) (16,0.1904) (17,0.255417) (18,0.203675) (19,0.207105) (20,0.23224) (21,0.153976) (22,0.193503) (23,0.202267) (24,0.128035) (25,0.20086) (26,0.224522) (27,0.197918) (28,0.218138) (29,0.195403) (30,0.206816) };
\addplot[clrDevice,mark=diamond, mark phase = 5] coordinates{(5,0.16791) (6,0.0630054) (7,0.0984934) (8,0.0614796) (9,0.0583313) (10,0.065946) (11,0.0924185) (12,0.00448941) (13,0.0640333) (14,0.0793263) (15,0.0439375) (16,0.0708472) (17,0.0412296) (18,0.0368644) (19,0.0580554) (20,0.0507088) (21,0.0685143) (22,0.0408822) (23,0.0411127) (24,0.0595599) (25,0.0349382) (26,0.0472644) (27,0.0261131) (28,0.046198) (29,0.0308349) (30,0.0524541) };
\addplot[clrZL,mark=pentagon, mark repeat=1] coordinates{(5,0.176036) (10,0.11763) (15,0.123031) (20,0.134368) };
\addplot[clrSN,mark=square, mark phase = 6] coordinates{(5,0.181667) (6,0.0886351) (7,0.156293) (8,0.126463) (9,0.123538) (10,0.155403) (11,0.183884) (12,0.131393) (13,0.16787) (14,0.166674) (15,0.169665) (16,0.194211) (17,0.171721) (18,0.160728) (19,0.168479) (20,0.172668) (21,0.178205) (22,0.185998) (23,0.170609) (24,0.166521) (25,0.1617) (26,0.173944) (27,0.169837) (28,0.196341) (29,0.18213) (30,0.19752) };
\addplot[clrSP,mark=triangle, mark phase = 6] coordinates{(5,0.183084) (6,0.100723) (7,0.167709) (8,0.145384) (9,0.174642) (10,0.188487) (11,0.225706) (12,0.191224) (13,0.211227) (14,0.225031) (15,0.251373) (16,0.229128) (17,0.229281) (18,0.20315) (19,0.228863) (20,0.231442) (21,0.223337) (22,0.224977) (23,0.229577) (24,0.210685) (25,0.240317) (26,0.233706) (27,0.246868) (28,0.256692) (29,0.231756) (30,0.255824) };

\end{axis}
\node[anchor=south west] at ($(current bounding box.north west)!.83!(current bounding box.north east)$) {\ref{milpimprlegend}};
\end{tikzpicture}

\begin{tikzpicture}[trim axis left, trim axis right]
\begin{axis}[xlabel={Number of tasks},ymax=500, height=0.23\textwidth,width=0.95\linewidth,legend to name=milptimelegend,legend cell align={left},legend columns = 3,legend style={font=\scriptsize},label style={font=\footnotesize},ylabel={Execution Time (ms)},every axis plot/.append style={thick,mark options={scale=0.5}, mark repeat=10}] 
\addlegendentry{WGDP\_Time}
\addlegendentry{WGDP\_Device}
\addlegendentry{ZhouLiu}
\addlegendentry{SingleNode}
\addlegendentry{SeriesParallel}

\addplot[clrTime,mark=*, mark phase = 4] coordinates{(5,19.4667) (6,23.4333) (7,29.6) (8,41) (9,38.5667) (10,80.8667) (11,92.9667) (12,85.8333) (13,82.8667) (14,106.733) (15,141.9) (16,157.433) (17,162.9) (18,174.467) (19,219.133) (20,265.567) (21,372.4) (22,407.533) (23,454.4) (24,460.667) (25,602.433) (26,719.6) };
\addplot[clrDevice,mark=diamond, mark phase = 5] coordinates{(5,12) (6,9.6) (7,12.7333) (8,20.7333) (9,24.1333) (10,63.0667) (11,38.2) (12,37.9333) (13,29.7667) (14,44.3) (15,72.0667) (16,50.1333) (17,54.9333) (18,55.6667) (19,64.8) (20,88.3333) (21,71.5333) (22,70.8667) (23,79.5333) (24,75.6333) (25,84.8667) (26,89.2667) (27,106.9) (28,101.567) (29,104.933) (30,107.867) };
\addplot[clrZL,mark=pentagon, mark repeat=1] coordinates{(5,67.5667) (10,1673.07) };
\addplot[clrSN,mark=square, mark phase = 13] coordinates{(5,0) (6,0) (7,0) (8,0) (9,0) (10,0.266667) (11,0.733333) (12,0.966667) (13,1.66667) (14,1.8) (15,2.66667) (16,3.76667) (17,4.06667) (18,4.56667) (19,6.16667) (20,6.33333) (21,8.36667) (22,10.1333) (23,9.83333) (24,10.9667) (25,12.6667) (26,18.1333) (27,20.5) (28,26.9667) (29,26.6) (30,34.6) };
\addplot[clrSP,mark=triangle, mark phase = 8] coordinates{(5,0) (6,0) (7,0) (8,0.133333) (9,0.466667) (10,0.766667) (11,1.43333) (12,1.53333) (13,2.7) (14,3.23333) (15,4) (16,6.13333) (17,6.96667) (18,7.1) (19,8.5) (20,10.3) (21,11.2333) (22,13.1) (23,14.5667) (24,13.2) (25,18.3333) (26,22.9667) (27,32.4667) (28,38.6333) (29,37.7) (30,40.2333) };
\end{axis}
\end{tikzpicture}
\caption{Comparison between single node and series-parallel decomposition mapping and three integer linear programs for random series-parallel graphs. Data points are generated for each graph size between \num{5} and \num{30} tasks for all algorithms except for \code{ZhouLiu}. Due to excessive execution times, data points for \code{ZhouLiu} are only available for \num{5}, \num{10}, \num{15} and \num{20} tasks.}
\label{fig:milpimprtime}
\end{figure}
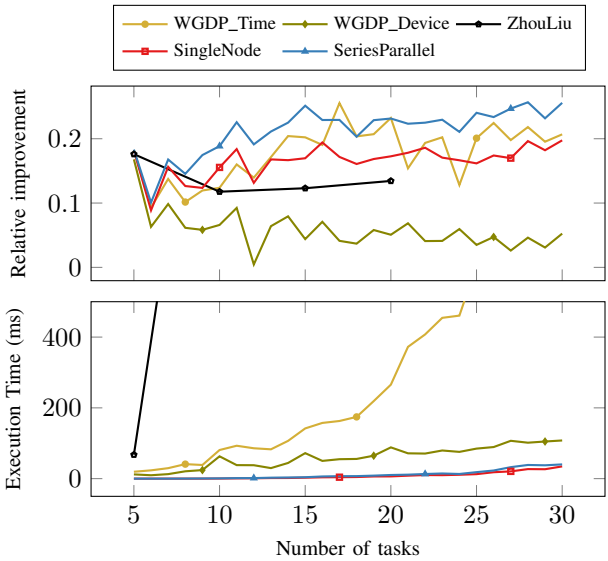

In Fig.~\ref{fig:milpimprtime}, the relative improvements and the respective execution times are depicted for the basic series-parallel and single node decomposition mapping algorithms as well as for the three reference MILPs. The MILP by Zhou and Liu timed out at a time limit of \num{5} minutes for graphs that have more than \num{20} nodes. It is able to find reasonably high improvements but can only be used for very small graphs due to a drastic increase in execution time. Generally, the time-based MILP leads to the highest improvements among the MILPs but also fails to always find the optimum. Its execution time strongly increases with increasing task size, taking about \num{1.5} seconds for \num{30} tasks and more than \num{16} minutes for \num{40} tasks. The single-node decomposition leads to a very fast mapping algorithm that produces good results with relative improvement values of \SIrange{10}{20}{\percent}. The series-parallel decomposition approach is able to improve on these results in exchange for a slightly higher execution time. Notably, it frequently finds better mappings than all of the considered MILPs as well. The only MILP that is comparable in terms of execution time is the device-based approach. However, this MILP leads to significantly smaller improvements than both decomposition mapping algorithms.

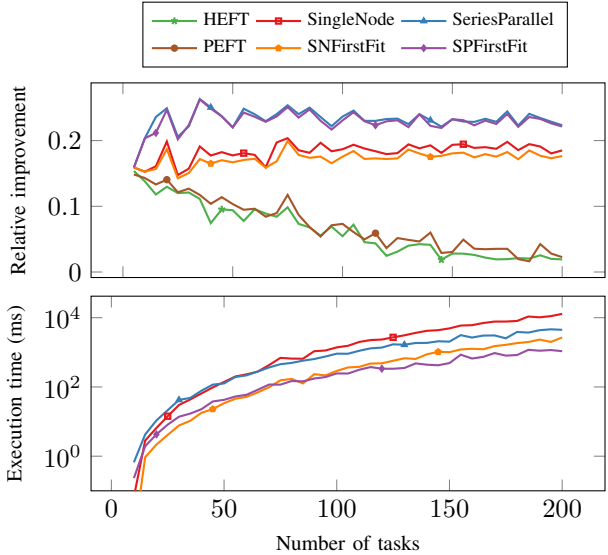
\begin{figure}[ht]
\centering
\begin{tikzpicture}[trim axis left, trim axis right]
\begin{axis}[height=0.23\textwidth,width=0.95\linewidth,legend to name=objlegend,legend cell align={left},legend columns = 3,legend style={font=\scriptsize},label style={font=\footnotesize},ylabel={Relative improvement},every axis plot/.append style={thick,mark options={scale=0.5}, mark repeat=20},xticklabel=\empty]
\addlegendentry{HEFT}
\addlegendentry{SingleNode}
\addlegendentry{SeriesParallel}
\addlegendentry{PEFT}
\addlegendentry{SNFirstFit}
\addlegendentry{SPFirstFit}

\addplot[clrHEFT,mark=star, mark options={scale=0.7}, mark phase = 9] coordinates{(5,0.153734) (10,0.137874) (15,0.117992) (20,0.129771) (25,0.120148) (30,0.121001) (35,0.111286) (40,0.0743589) (45,0.0953618) (50,0.0940166) (55,0.0776774) (60,0.0956567) (65,0.0889438) (70,0.0841476) (75,0.0984543) (80,0.0733414) (85,0.0678343) (90,0.0553789) (95,0.0691922) (100,0.0544374) (105,0.0720901) (110,0.0453903) (115,0.043634) (120,0.0246216) (125,0.0310488) (130,0.0400767) (135,0.0423722) (140,0.0413617) (145,0.0187523) (150,0.0281016) (155,0.0279507) (160,0.026006) (165,0.0219284) (170,0.0192298) (175,0.0195625) (180,0.0211507) (185,0.0204541) (190,0.0254759) (195,0.0200402) (200,0.0192397) };

\addplot[clrSN,mark=square, mark phase = 11] coordinates{(5,0.159072) (10,0.152601) (15,0.160914) (20,0.198632) (25,0.147465) (30,0.157314) (35,0.191108) (40,0.177277) (45,0.182407) (50,0.177457) (55,0.180883) (60,0.178198) (65,0.159289) (70,0.197185) (75,0.203653) (80,0.185283) (85,0.181437) (90,0.196654) (95,0.183751) (100,0.186847) (105,0.193708) (110,0.187926) (115,0.183794) (120,0.179451) (125,0.180905) (130,0.194257) (135,0.188174) (140,0.192761) (145,0.181301) (150,0.19357) (155,0.19446) (160,0.188847) (165,0.189996) (170,0.187487) (175,0.198058) (180,0.185475) (185,0.194606) (190,0.190988) (195,0.180478) (200,0.185065) };

\addplot[clrSP,mark=triangle, mark phase = 8] coordinates{(5,0.159072) (10,0.203754) (15,0.23586) (20,0.249185) (25,0.205724) (30,0.221748) (35,0.263127) (40,0.250461) (45,0.236982) (50,0.220262) (55,0.248511) (60,0.240102) (65,0.229503) (70,0.240254) (75,0.253768) (80,0.240191) (85,0.250159) (90,0.235878) (95,0.221851) (100,0.236518) (105,0.24563) (110,0.230032) (115,0.23003) (120,0.232822) (125,0.233393) (130,0.225061) (135,0.24011) (140,0.230938) (145,0.221172) (150,0.232448) (155,0.229007) (160,0.228462) (165,0.233218) (170,0.228547) (175,0.244396) (180,0.222952) (185,0.240722) (190,0.234443) (195,0.228737) (200,0.223354) };

\addplot[clrPEFT,mark=*, mark phase = 4, mark repeat=19] coordinates{(5,0.148443) (10,0.143) (15,0.133248) (20,0.140517) (25,0.121411) (30,0.126864) (35,0.117302) (40,0.103748) (45,0.113913) (50,0.103398) (55,0.0947165) (60,0.0962892) (65,0.0841774) (70,0.089628) (75,0.117353) (80,0.0869777) (85,0.0683807) (90,0.0539684) (95,0.0711232) (100,0.073274) (105,0.060804) (110,0.049673) (115,0.0590983) (120,0.036408) (125,0.0516501) (130,0.0488989) (135,0.0464642) (140,0.0600269) (145,0.0290277) (150,0.0305051) (155,0.0492501) (160,0.0354459) (165,0.0346975) (170,0.0352808) (175,0.0353083) (180,0.0196793) (185,0.0163077) (190,0.04247) (195,0.0280982) (200,0.0226554) };

\addplot[clrSNFF,mark=pentagon, mark phase = 8] coordinates{(5,0.159072) (10,0.152661) (15,0.156781) (20,0.186812) (25,0.14269) (30,0.150687) (35,0.171895) (40,0.164982) (45,0.170175) (50,0.166904) (55,0.170508) (60,0.172406) (65,0.158826) (70,0.168473) (75,0.199342) (80,0.178137) (85,0.173796) (90,0.175651) (95,0.165548) (100,0.175476) (105,0.18412) (110,0.172065) (115,0.172888) (120,0.172036) (125,0.172874) (130,0.18634) (135,0.180387) (140,0.175233) (145,0.176786) (150,0.180532) (155,0.181639) (160,0.174525) (165,0.179397) (170,0.17558) (175,0.18263) (180,0.171656) (185,0.185017) (190,0.177028) (195,0.173128) (200,0.176549) };

\addplot[clrSPFF,mark=diamond, mark phase = 3] coordinates{(5,0.159072) (10,0.203558) (15,0.211722) (20,0.246868) (25,0.201925) (30,0.223709) (35,0.262556) (40,0.248629) (45,0.236927) (50,0.219905) (55,0.242669) (60,0.236188) (65,0.228595) (70,0.23673) (75,0.250977) (80,0.234981) (85,0.247378) (90,0.230134) (95,0.216795) (100,0.230676) (105,0.243206) (110,0.229451) (115,0.223378) (120,0.229323) (125,0.230957) (130,0.220482) (135,0.240095) (140,0.222479) (145,0.219193) (150,0.232586) (155,0.231056) (160,0.223218) (165,0.23044) (170,0.22528) (175,0.240431) (180,0.220963) (185,0.235882) (190,0.232989) (195,0.226213) (200,0.221452) };

\end{axis}
\node[anchor=south west] at ($(current bounding box.north west)!.80!(current bounding box.north east)$) {\ref{objlegend}};
\end{tikzpicture}

\begin{tikzpicture}[trim axis left, trim axis right]
\begin{axis}[xlabel={Number of tasks}, ymin=0.1, height=0.23\textwidth,width=0.95\linewidth,legend to name=execlegend,legend cell align={left},legend columns = 4,legend style={font=\scriptsize},label style={font=\footnotesize},ylabel={Execution time (ms)},ymode=log,every axis plot/.append style={thick,mark options={scale=0.5}, mark repeat=20}]

\addlegendentry{SingleNode}
\addlegendentry{SNFirstFit}
\addlegendentry{SeriesParallel}
\addlegendentry{SPFirstFit}

\addplot[clrSN,mark=square, mark phase = 3] coordinates{(5,0) (10,0.0666667) (15,2.83333) (20,6.53333) (25,14.2667) (30,30.1667) (35,43.4667) (40,65.1667) (45,96.8667) (50,139.3) (55,200.233) (60,231.4) (65,278.667) (70,429.267) (75,691.867) (80,658.9) (85,653.633) (90,1077.8) (95,1130.33) (100,1390.53) (105,1545.93) (110,1997.4) (115,2259.4) (120,2343) (125,2713.87) (130,3099.8) (135,3651.33) (140,4186.53) (145,4366.57) (150,4914.3) (155,5933.2) (160,6086.7) (165,7023.53) (170,7735.93) (175,7743.73) (180,8061.17) (185,10818.8) (190,10264.3) (195,11118.6) (200,12906.1) };

\addplot[clrSNFF,mark=pentagon, mark phase = 7] coordinates{(5,0) (10,0.004) (15,0.933333) (20,2.16667) (25,4.13333) (30,7.7) (35,10.5333) (40,17.9333) (45,22.9667) (50,34) (55,45.5667) (60,51.7333) (65,69.8333) (70,98.5333) (75,155.1) (80,170.967) (85,126.433) (90,233.467) (95,215.967) (100,289.933) (105,366.1) (110,381.7) (115,473.8) (120,483.1) (125,570.467) (130,671.933) (135,644.867) (140,877.7) (145,1030.97) (150,1011.2) (155,1209.37) (160,1267.47) (165,1229.1) (170,1511.17) (175,1645.73) (180,1852) (185,1983.67) (190,2331.03) (195,2001.7) (200,2683.07) };

\addplot[clrSP,mark=triangle, mark phase = 5] coordinates{(5,0) (10,0.666667) (15,4.23333) (20,10.8) (25,21.3) (30,42.2333) (35,48.2) (40,78.1667) (45,116.633) (50,126.467) (55,194.1) (60,216.467) (65,279.333) (70,366.467) (75,454.067) (80,495.1) (85,574.867) (90,640.9) (95,757.367) (100,914) (105,910.133) (110,1110.9) (115,1303.67) (120,1372.73) (125,1701.33) (130,1656.83) (135,1861.57) (140,1873) (145,2084.3) (150,2036.8) (155,3136.57) (160,2674.1) (165,3048.6) (170,3074.27) (175,2541.23) (180,3876.13) (185,3722.83) (190,4354.43) (195,4622.53) (200,4468.97) };

\addplot[clrSPFF,mark=diamond, mark phase = 3] coordinates{(5,0) (10,0.233333) (15,1.93333) (20,4.26667) (25,8.16667) (30,13.8) (35,17) (40,22.8333) (45,38.1) (50,42.4667) (55,52.9333) (60,59.4333) (65,83.1333) (70,117.267) (75,116.667) (80,148.233) (85,145.067) (90,175) (95,193.233) (100,246.033) (105,244.167) (110,314.333) (115,374.967) (120,336.9) (125,335.5) (130,353.033) (135,478.767) (140,439.4) (145,422.967) (150,488.233) (155,854.6) (160,662.567) (165,741.167) (170,952.1) (175,804.167) (180,835.067) (185,1179.53) (190,1113.83) (195,1163.87) (200,1078.9) };

\end{axis}
\end{tikzpicture}
\caption{Comparison of the relative improvements and the execution times of the list-based scheduling algorithms \code{HEFT} and \code{PEFT} and the two decomposition strategies \code{SingleNode} and \code{SeriesParallel} with and without the FirstFit heuristic. Data points are generated for \num{5} to {200} tasks with steps of \num{5} tasks. The execution time is displayed using a logarithmic scale. The execution times for \code{HEFT} and \code{PEFT} are below \SI{10}{\micro\second} and therefore not displayed.}
\label{fig:heftimprtime}
\end{figure}

The HEFT and PEFT algorithms can be applied to significantly larger task graphs than the MILPs. In Fig.~\ref{fig:heftimprtime} the relative improvement of HEFT and PEFT is displayed in relation to the improvement achieved by the two decomposition mapping algorithms in both the basic and the FirstFit variant (cf.\ Sec.~\ref{ssc:optimizations}). Both HEFT and PEFT took less than \SI{10}{\micro\second} to run for all depicted graph sizes. While for small graphs all algorithms show similar improvements, the quality of the mapping produced by both HEFT and PEFT decreases with increasing graph size. This effect can be observed not only for the relative, but also for the absolute improvement. For a small number of tasks, an improved mapping can be found for almost all generated graphs through both algorithms, whereas for a large number of tasks HEFT can only find an improved mapping for about \SI{50}{\percent} and PEFT for about \SI{70}{\percent} of the graphs. This is a direct consequence of the local view both algorithms obtain during the scheduling phase, where they can only very roughly estimate the impact of the mapping of a single task on the overall execution. In contrast, the relative improvement achieved by the decomposition mapping algorithms stays roughly the same, with the series-parallel decomposition finding on average an about \SI{5}{\percent} higher improvement.

Notably, the difference in the achieved makespan between the basic decomposition mapping principle and the FirstFit heuristic is almost negligible. At the same time, the FirstFit heuristic is able to reduce the execution time for large graphs by up to \SI{75}{\percent} for the series-parallel decomposition and up to \SI{80}{\percent} for the single node decomposition. This suggests that in most cases it is sufficient to use FirstFit instead of the significantly more expensive basic variant. In particular, in our test cases, using a $\gamma$-threshold heuristic with $\gamma>1$ does not provide a significant benefit in comparison with the FirstFit variant of the mapping algorithm.

Generally, on our test data, all decomposition-based mapping strategies exhibit a quadratic behavior regarding their execution time, although their theoretical execution time has a cubic dependency on the number of tasks (cf.\ Sec.~\ref{ssc:decomp_principle}). This is the case since the number of iterations in which an improvement occurs is in practice much smaller than the number of tasks and grows very slowly. Especially for the series-parallel decomposition, a larger graph often means that larger subgraphs can be replaced at once, effectively reducing the number of necessary iterations. In consequence, for large graphs the series-parallel decomposition mapping can be computed faster than the single node decomposition mapping. As depicted in Fig.~\ref{fig:heftimprtime}, the series-parallel decomposition leads to a lower execution time starting at about \num{50} tasks for the basic principle and at about \num{75} tasks for the FirstFit heuristic.

\begin{figure}[ht]
\centering
\begin{tikzpicture}[trim axis left, trim axis right]
\begin{axis}[height=0.23\textwidth,width=0.95\linewidth,legend to name=metalegend,legend cell align={left},legend columns = 4,legend style={font=\scriptsize},label style={font=\footnotesize},ylabel={Relative improvement},every axis plot/.append style={thick,mark options={scale=0.5}, mark repeat=5},xticklabel=\empty]
\addlegendentry{SNFirstFit}
\addlegendentry{SPFirstFit}
\addlegendentry{NSGAII}

\addplot[clrSNFF,mark=pentagon, mark phase = 3] coordinates{(5,0.127337) (10,0.150046) (15,0.170741) (20,0.168501) (25,0.169327) (30,0.139381) (35,0.168753) (40,0.184906) (45,0.185251) (50,0.187355) (55,0.16637) (60,0.16373) (65,0.184313) (70,0.180891) (75,0.171069) (80,0.177626) (85,0.181872) (90,0.181406) (95,0.167864) (100,0.180817) };
\addplot[clrSPFF,mark=diamond, mark phase = 3] coordinates{(5,0.132547) (10,0.191759) (15,0.239507) (20,0.218425) (25,0.23736) (30,0.231809) (35,0.232844) (40,0.274376) (45,0.25273) (50,0.242761) (55,0.243844) (60,0.24406) (65,0.239383) (70,0.25136) (75,0.234845) (80,0.246482) (85,0.231671) (90,0.244342) (95,0.222201) (100,0.238339) };
\addplot[clrNSGA, mark=square, mark phase = 5] coordinates{(5,0.132547) (10,0.196099) (15,0.246049) (20,0.197358) (25,0.220297) (30,0.20029) (35,0.200692) (40,0.239873) (45,0.204399) (50,0.228876) (55,0.214863) (60,0.205786) (65,0.212342) (70,0.212434) (75,0.203812) (80,0.213492) (85,0.211851) (90,0.214598) (95,0.201806) (100,0.208318) };

\end{axis}
\node[anchor=south west] at ($(current bounding box.north west)!.84!(current bounding box.north east)$) {\ref{metalegend}};
\end{tikzpicture}

\begin{tikzpicture}[trim axis left, trim axis right]
\begin{axis}[xlabel={Number of tasks}, ymin=1, height=0.23\textwidth,width=0.95\linewidth,legend to name=metaexeclegend,legend cell align={left},legend columns = 4,legend style={font=\scriptsize},label style={font=\footnotesize},ylabel={Execution time (ms)},ymode=log,every axis plot/.append style={thick,mark options={scale=0.5}, mark repeat=5}]

\addlegendentry{SNFirstFit}
\addlegendentry{SPFirstFit}
\addlegendentry{NSGAII}

\addplot[clrSNFF,mark=pentagon, mark phase = 7] coordinates{(5,0) (10,0.1) (15,0.8) (20,1.56667) (25,5.26667) (30,5.93333) (35,11.4333) (40,20.7333) (45,24.2333) (50,38.2333) (55,54.9667) (60,61.4333) (65,87.6667) (70,106.067) (75,114.467) (80,142.9) (85,207.767) (90,210.6) (95,241.167) (100,410.7) };

\addplot[clrSPFF,mark=diamond, mark phase = 3] coordinates{(5,37.3667) (10,41.7667) (15,46.2333) (20,51.6333) (25,59.5333) (30,68.2) (35,78.5333) (40,98.6667) (45,106.667) (50,123.267) (55,128.833) (60,146.833) (65,192.967) (70,211.967) (75,213.567) (80,224.3) (85,284.033) (90,316.4) (95,300.733) (100,365.433) };
\addplot[clrNSGA, mark=square, mark phase = 5] coordinates{(5,258.333) (10,553.867) (15,1000.17) (20,1153.1) (25,1746.2) (30,2098.7) (35,2415.23) (40,3397.57) (45,3594.13) (50,4107.77) (55,5362.7) (60,5873.9) (65,5958.1) (70,6586.47) (75,7423.17) (80,8145.43) (85,8933.93) (90,9378.6) (95,9726.33) (100,11265.8) };

\end{axis}
\end{tikzpicture}
\caption{Comparison of the relative improvement and the execution times of the genetic algorithm (\code{NSGAII}) and the two decomposition strategies with the FirstFit heuristic. Data points are generated for \num{5} to {100} tasks with steps of \num{5} tasks. The execution time is displayed using a logarithmic scale.}
\label{fig:metaimprtime}
\end{figure}
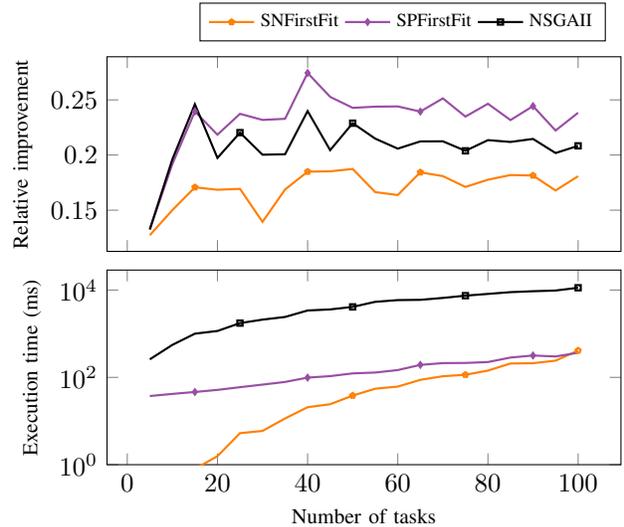

\begin{figure}[ht]
\centering
\begin{tikzpicture}[trim axis left, trim axis right]
\begin{axis}[height=0.23\textwidth,width=0.95\linewidth,legend to name=geneticlegend,legend cell align={left},legend columns = 4,legend style={font=\scriptsize},label style={font=\footnotesize},ylabel={Relative improvement},every axis plot/.append style={thick,mark options={scale=0.5}, mark repeat=5},xticklabel=\empty]
\addlegendentry{SNFirstFit}
\addlegendentry{SPFirstFit}
\addlegendentry{NSGAII}

\addplot[clrSNFF,mark=pentagon, mark phase = 4] coordinates{(50,0.176032) (100,0.176032) (150,0.176032) (200,0.176032) (250,0.176032) (300,0.176032) (350,0.176032) (400,0.176032) (450,0.176032) (500,0.176032) };
\addplot[clrSPFF,mark=diamond, mark phase = 3] coordinates{(50,0.22545) (100,0.22545) (150,0.22545) (200,0.22545) (250,0.22545) (300,0.22545) (350,0.22545) (400,0.22545) (450,0.22545) (500,0.22545) };
\addplot[clrNSGA, mark=square, mark phase = 2] coordinates{(50,0.110533) (100,0.157835) (150,0.18326) (200,0.189629) (250,0.193293) (300,0.192195) (350,0.19039) (400,0.195818) (450,0.1888) (500,0.196342) };

\end{axis}
\node[anchor=south west] at ($(current bounding box.north west)!.78!(current bounding box.north east)$) {\ref{geneticlegend}};
\end{tikzpicture}

\begin{tikzpicture}[trim axis left, trim axis right]
\begin{axis}[xlabel={Number of generations}, height=0.23\textwidth,width=0.95\linewidth,legend to name=geneticexeclegend,legend cell align={left},legend columns = 4,legend style={font=\scriptsize},label style={font=\footnotesize},ylabel={Execution time (ms)},every axis plot/.append style={thick,mark options={scale=0.5}, mark repeat=5},ymode=log]

\addlegendentry{SNFirstFit}
\addlegendentry{SPFirstFit}
\addlegendentry{NSGAII}

\addplot[clrSNFF,mark=pentagon, mark phase = 4] coordinates{(50,2318.1) (100,2318.1) (150,2318.1) (200,2318.1) (250,2318.1) (300,2318.1) (350,2318.1) (400,2318.1) (450,2318.1) (500,2318.1) };
\addplot[clrSPFF,mark=diamond, mark phase = 3] coordinates{(50,1222.37) (100,1222.37) (150,1222.37) (200,1222.37) (250,1222.37) (300,1222.37) (350,1222.37) (400,1222.37) (450,1222.37) (500,1222.37) };
\addplot[clrNSGA, mark=square, mark phase = 2] coordinates{(50,2561.17) (100,5311.67) (150,8433.47) (200,11134.4) (250,13778.3) (300,17993.9) (350,19668.3) (400,23588.2) (450,26432.3) (500,30114.4) };

\end{axis}
\end{tikzpicture}
\caption{Tradeoff between execution time and relative improvement for the NSGA-II genetic algorithm. Data points are generated for \num{50} to \num{500} generations with steps of \num{50} generations and represent the average over \num{30} random series-parallel graphs with \num{200} nodes. The execution time is displayed using a logarithmic scale.}
\label{fig:generations}
\end{figure}
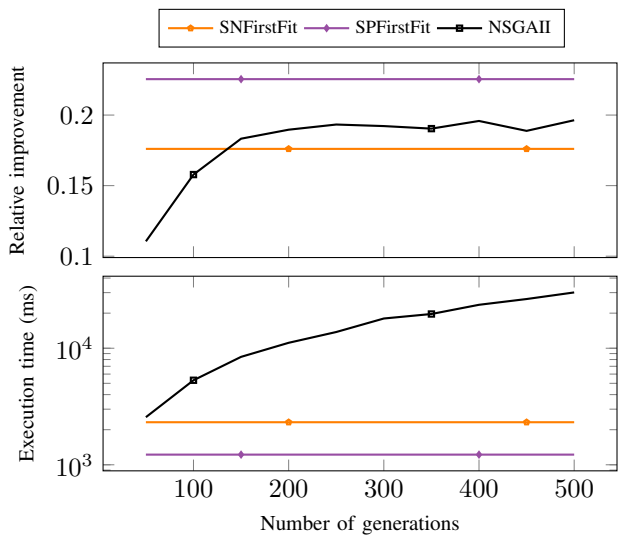

The behavior of the genetic algorithm is shown in Fig.~\ref{fig:metaimprtime}. It is able to reliably deal with local minima, resulting in a high relative improvement independent of the size of the task graph. While it can often find slightly better results than the single node decomposition approach, it is still frequently outperformed by the series-parallel decomposition mapping.
The execution time of the genetic algorithm strongly increases with the size of the task graph At $n=100$, the metaheuristic is \num{30} times slower than the decomposition mapping approaches. For the evaluation of individuals, we use the same model-based evaluation function as for the decomposition-based approaches in order to ensure fairness. While the computation of this function is slightly more expensive (around factor~\num{3} compared to the maximum sum of computation times of all tasks over all devices), simpler cost functions lead to a worse capturing of the system behavior and, hence, to a strongly reduced mapping quality up to the point that the metaheuristic is not able to find any actual improvement at all. 

Aside from the cost function, the execution time of the genetic algorithm strongly depends on the used parameterization. In particular, it may be accelerated through a better monitoring of the number of generations needed until reaching a saturation~\cite{ravber2022}. In Fig.~\ref{fig:generations}, the behavior of NSGA-II is shown for various numbers of generations on random series-parallel graphs of size \num{200}. In this case, the saturation starts at about \num{200} generations, which reduces the execution time by \SI{63}{\percent} compared to an evaluation with \num{500} generations. Nevertheless, the execution time at this point is still \num{5} to \num{10} times as high as the execution time of the single node and series-parallel decomposition approach, respectively. Note that with a constant amount of generations, the genetic algorithm grows slower than the decomposition approaches. However, not adjusting the generations to the graph size leads to a deterioration of the result quality. At a graph size of \num{500} nodes, the execution time of NSGA-II with \num{200} generations is comparable to the execution time of the single node approach and twice as large as the one of the series-parallel approach, but finds only a \SI{12}{\percent} relative improvement compared to a \SI{17}{\percent} improvement found by the decomposition-based heuristics.

\subsection{Random almost series-parallel graphs}

In the previous section, we considered graphs that are guaranteed to be series-parallel and therefore have a unique series-parallel decomposition tree. In practice, one may encounter task graphs, which are \emph{almost series-parallel}, i.e., graphs with a mostly series-parallel structure but a few conflicting edges. We generate almost series-parallel graphs by generating a series-parallel graph with the desired number of nodes and randomly inserting $k$ new edges, which are directed according to a random topological order. Since in a series-parallel graph there can only be a linear number of non-conflicting edges, most of the newly generated edges will be conflicting.

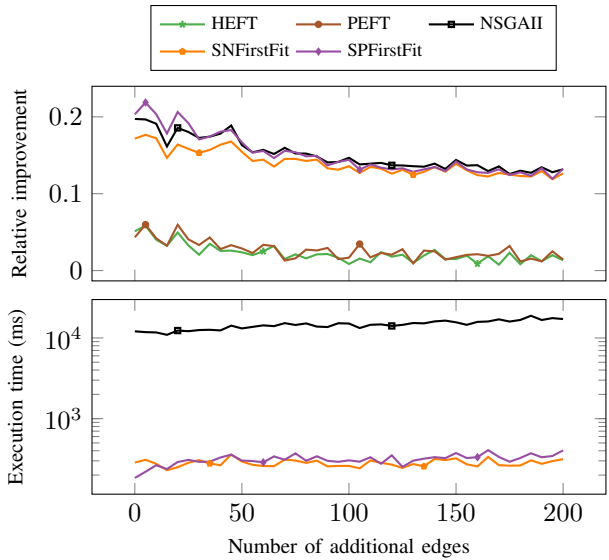
\begin{figure}[ht]
\centering
\begin{tikzpicture}[trim axis left, trim axis right]
\begin{axis}[height=0.23\textwidth,width=0.95\linewidth,legend to name=almostsplegend,legend cell align={left},legend columns = 3,legend style={font=\scriptsize},label style={font=\footnotesize},ylabel={Relative improvement},every axis plot/.append style={thick,mark options={scale=0.5}, mark repeat=20},xticklabel=\empty]
\addlegendentry{HEFT}
\addlegendentry{PEFT}
\addlegendentry{NSGAII}
\addlegendentry{SNFirstFit}
\addlegendentry{SPFirstFit}

\addplot[clrHEFT,mark=star, mark options={scale=0.7}, mark phase = 13] coordinates{(0,0.0510947) (5,0.0585082) (10,0.040246) (15,0.0323306) (20,0.0495632) (25,0.0327401) (30,0.0206661) (35,0.0348674) (40,0.0255967) (45,0.0259985) (50,0.0238068) (55,0.0201056) (60,0.025123) (65,0.032201) (70,0.0150998) (75,0.0211029) (80,0.0159378) (85,0.0211235) (90,0.0215257) (95,0.016799) (100,0.00856667) (105,0.0156761) (110,0.0108695) (115,0.0240179) (120,0.0181717) (125,0.0206845) (130,0.00996726) (135,0.0195313) (140,0.0269057) (145,0.0149989) (150,0.015082) (155,0.0194532) (160,0.00899004) (165,0.01866) (170,0.00771161) (175,0.0232506) (180,0.00824306) (185,0.0200651) (190,0.0120758) (195,0.0200851) (200,0.0133981) };

\addplot[clrPEFT,mark=*, mark phase = 2] coordinates{(0,0.0433422) (5,0.0597862) (10,0.0420072) (15,0.0323768) (20,0.0595983) (25,0.0406665) (30,0.0332441) (35,0.0430652) (40,0.0280287) (45,0.0330553) (50,0.0287684) (55,0.0227959) (60,0.0335829) (65,0.0317828) (70,0.0130752) (75,0.015939) (80,0.027303) (85,0.0261939) (90,0.0294916) (95,0.0149137) (100,0.0168283) (105,0.0345985) (110,0.0172618) (115,0.022983) (120,0.0206906) (125,0.0278612) (130,0.00924438) (135,0.026054) (140,0.0250689) (145,0.0143001) (150,0.0174902) (155,0.0204321) (160,0.0213132) (165,0.0191765) (170,0.0217387) (175,0.0320796) (180,0.0118618) (185,0.0155984) (190,0.0120534) (195,0.0250271) (200,0.0144176) };

\addplot[clrNSGA, mark=square, mark phase = 5] coordinates{(0,0.197288) (5,0.196633) (10,0.191011) (15,0.161475) (20,0.18566) (25,0.180324) (30,0.172592) (35,0.174255) (40,0.178176) (45,0.188751) (50,0.16321) (55,0.153819) (60,0.157038) (65,0.151673) (70,0.159733) (75,0.152321) (80,0.152216) (85,0.148452) (90,0.140721) (95,0.141326) (100,0.146766) (105,0.138111) (110,0.139108) (115,0.140107) (120,0.136978) (125,0.136743) (130,0.135799) (135,0.135166) (140,0.139161) (145,0.132195) (150,0.143916) (155,0.13664) (160,0.137122) (165,0.129389) (170,0.135576) (175,0.125642) (180,0.129893) (185,0.127218) (190,0.134426) (195,0.127994) (200,0.131735) };

\addplot[clrSNFF,mark=pentagon, mark phase = 7] coordinates{(0,0.17166) (5,0.17664) (10,0.172114) (15,0.146649) (20,0.16399) (25,0.158385) (30,0.153151) (35,0.157089) (40,0.16377) (45,0.16793) (50,0.154495) (55,0.142587) (60,0.144321) (65,0.135275) (70,0.145279) (75,0.145305) (80,0.142617) (85,0.144361) (90,0.133033) (95,0.131283) (100,0.135799) (105,0.126928) (110,0.134957) (115,0.132549) (120,0.126101) (125,0.131069) (130,0.124668) (135,0.128462) (140,0.134704) (145,0.12869) (150,0.138993) (155,0.130505) (160,0.124242) (165,0.122433) (170,0.127067) (175,0.124911) (180,0.123122) (185,0.122392) (190,0.129452) (195,0.118891) (200,0.126437) };

\addplot[clrSPFF,mark=diamond, mark phase = 2] coordinates{(0,0.203282) (5,0.218696) (10,0.203454) (15,0.178392) (20,0.2063) (25,0.191897) (30,0.170662) (35,0.174413) (40,0.180738) (45,0.182767) (50,0.166784) (55,0.153451) (60,0.155667) (65,0.14636) (70,0.155797) (75,0.153984) (80,0.148682) (85,0.149377) (90,0.136857) (95,0.141423) (100,0.144028) (105,0.131346) (110,0.138051) (115,0.13385) (120,0.131604) (125,0.133084) (130,0.128619) (135,0.131677) (140,0.134718) (145,0.129512) (150,0.14198) (155,0.13164) (160,0.127987) (165,0.127041) (170,0.131751) (175,0.124252) (180,0.127892) (185,0.123232) (190,0.133191) (195,0.119405) (200,0.132447) };

\end{axis}
\node[anchor=south west] at ($(current bounding box.north west)!.78!(current bounding box.north east)$) {\ref{almostsplegend}};
\end{tikzpicture}

\begin{tikzpicture}[trim axis left, trim axis right]
\begin{axis}[xlabel={Number of additional edges}, height=0.23\textwidth,width=0.95\linewidth,legend to name=almostspexeclegend,legend cell align={left},legend columns = 4,legend style={font=\scriptsize},label style={font=\footnotesize},ylabel={Execution time (ms)},every axis plot/.append style={thick,mark options={scale=0.5}, mark repeat=20}, ymode=log]

\addlegendentry{NSGAII}
\addlegendentry{SNFirstFit}
\addlegendentry{SPFirstFit}

\addplot[clrNSGA, mark=square, mark phase = 5] coordinates{(0,12075.7) (5,11799.8) (10,11699.7) (15,10932.4) (20,12341.7) (25,12119.6) (30,12524) (35,12617.9) (40,12389.1) (45,14203.1) (50,13117.6) (55,13713.8) (60,14312.4) (65,14005) (70,15252.1) (75,14479.1) (80,15134.7) (85,13822.9) (90,13652.2) (95,15200.7) (100,15085.9) (105,13259.4) (110,14547.2) (115,14715.5) (120,14089.3) (125,14521.8) (130,15309.7) (135,15150.6) (140,16023.6) (145,16384.8) (150,15621.6) (155,14536.2) (160,15810) (165,15998.3) (170,16994.7) (175,15964) (180,16699.1) (185,18824.3) (190,16686.8) (195,17606.1) (200,17138) };
\addplot[clrSNFF,mark=pentagon, mark phase = 8] coordinates{(0,286.067) (5,309.333) (10,275.667) (15,229.833) (20,250.333) (25,284.167) (30,306.6) (35,278.5) (40,264.6) (45,359.733) (50,294.167) (55,267.867) (60,258.6) (65,258.1) (70,310.3) (75,302.7) (80,283.3) (85,301.7) (90,256.3) (95,259.6) (100,259.567) (105,242.633) (110,303.3) (115,283.6) (120,269.8) (125,245.2) (130,274.167) (135,256.6) (140,317.3) (145,306.367) (150,322.7) (155,271.767) (160,256.133) (165,336.233) (170,265) (175,261.633) (180,262.9) (185,304.667) (190,275.667) (195,297.767) (200,314.933) };
\addplot[clrSPFF,mark=diamond, mark phase = 13] coordinates{(0,184.233) (5,219.867) (10,264.867) (15,237.433) (20,290.167) (25,308.433) (30,292.067) (35,292.8) (40,330.5) (45,357.5) (50,302.533) (55,298.733) (60,287.633) (65,340.967) (70,309.667) (75,371.767) (80,300.833) (85,341.8) (90,300.267) (95,292.033) (100,304.567) (105,293.033) (110,332.333) (115,276.333) (120,351.367) (125,252.4) (130,302.033) (135,319.133) (140,333.267) (145,324.567) (150,375.3) (155,325.333) (160,332.933) (165,407.633) (170,335.533) (175,292.7) (180,325.433) (185,372.733) (190,332.633) (195,344.433) (200,404.833) };

\end{axis}
\end{tikzpicture}
\caption{Comparison between \code{HEFT}, \code{PEFT}, \code{NSGAII} and the single node and series-parallel decomposition strategies with the FirstFit heuristic for task graphs with \num{100} nodes and an increasing number of potentially conflicting edges. Data points are generated for \num{5} to {200} additional edges with steps of \num{5} edges. The execution times for \code{HEFT} and \code{PEFT} are below \SI{10}{\micro\second} and therefore not displayed.}
\label{fig:conflicting}
\end{figure}

Fig.~\ref{fig:conflicting} shows the relative improvement obtained using the decomposition mapping and the list scheduling strategies for task graphs with \num{100} nodes and between \num{0} and \num{200} additional edges. The added complexity due to the increased data transfer leads to a slight decrease in quality for all algorithms. With an increasing number of additional edges, the series-parallel decomposition converges towards a single node decomposition as all decomposition trees converge towards single edges. With a high number of additional edges, the relative improvements achieved by the genetic algorithm closely resemble the improvements achieved by the decomposition-based heuristics. This observation suggests that with an increasing number of edges the maximum (optimal) gain decreases and the mapping found by all three algorithms is still close to optimal.
Adding up to \num{200} random edges has no significant effect on the execution time of the single node decomposition algorithm. The execution time of the series-parallel decomposition algorithm, however, increases as large series-parallel subgraphs are split into many smaller subgraphs. At about \num{15} additional edges, the series-parallel decomposition approach becomes slightly slower than the single node decomposition approach. With \num{200} additional edges, the execution time is about \SI{30}{\percent} higher than the execution time of the single node approach.

\subsection{Real-world examples}
\label{ssc:realworld}

Real-world benchmarks for comprehensive, strongly heterogeneous systems are hard to obtain. Complex real-world systems, which consist of many different tasks and make use of a variety of different processing units, are usually not accessible for the public. 
The WfCommons project~\cite{coleman2022} aims to create an open-source collection of realistic workflow generators based on real-world applications, such as the well-known montage~\cite{berriman2004} and epigenomics~\cite{juve2013} datasets. Workflows from this dataset have been employed to evaluate heterogeneous scheduling algorithms~\cite{maurya2018, sukhoroslov2023}, albeit only with respect to heterogeneous CPU clusters without GPUs or FPGAs. 
Based on the WfCommons project, Sukhoroslov and Gorokhovskii created a fixed benchmark set for the evaluation of heterogeneous DAG scheduling algorithms~\cite{sukhoroslov2023}. The benchmark set consists of \num{150} graphs of different graph sizes divided into nine different graph types. The largest graphs occur in the \code{montage} and \code{epigenomics} datasets, which contain workflows with up to \num{1312} and up to \num{1695} tasks, respectively. In this section, we evaluate the decomposition mapping principle on test cases derived from their benchmark instances. For this, we recreate the task graphs, the input and output data sizes as well as the task complexity from the provided workflows and augment these tasks by random parallelizability and streamability values analogously to Section~\ref{ssc:seriesparallel}.

\begin{table}
\centering
\caption{Relative improvements and total execution times for the benchmark sets from~\cite{sukhoroslov2023}. For each set, the first row denotes the average positive relative improvement among all graphs of the set. The second row shows the summed up execution time of all graphs from the respective benchmark set, where the execution time for each graph is averaged over \num{10} runs with different parameterizations.}
\begin{tabular}{c|ccccc}
								& HEFT					& PEFT						& NSGAII					& SNFirstFit				& SPFirstFit \\
\hline\hline
\multirow{2}{*}{1000genome} 		& \SI{5}{\percent} 		& \SI{7}{\percent} 			& \SI{9}{\percent} 		& \SI{9}{\percent} 		& \SI{9}{\percent} \\
								& \SI{4}{\milli\second}	& \SI{8}{\milli\second} 		& \SI{296}{\second} 		& \SI{5}{\second}		& \SI{6}{\second} \\
								\hline
\multirow{2}{*}{blast} 			& \SI{0}{\percent} 		& \SI{0}{\percent} 			& \SI{8}{\percent} 		& \SI{7}{\percent} 		& \SI{7}{\percent} \\
								& \SI{0}{\milli\second}	& \SI{0}{\milli\second} 		& \SI{92}{\second} 		& \SI{1}{\second}		& \SI{1}{\second} \\
								\hline
\multirow{2}{*}{cycles} 			& \SI{0}{\percent} 		& \SI{0}{\percent} 			& \SI{12}{\percent} 		& \SI{12}{\percent} 		& \SI{11}{\percent} \\
								& \SI{7}{\milli\second}	& \SI{12}{\milli\second} 	& \SI{257}{\second} 		& \SI{10}{\second}		& \SI{27}{\second} \\
								\hline
\multirow{2}{*}{epigenomics} 	& \SI{2}{\percent} 		& \SI{2}{\percent} 			& \SI{20}{\percent} 		& \SI{20}{\percent} 		& \SI{20}{\percent} \\
								& \SI{15}{\milli\second}	& \SI{32}{\milli\second} 	& \SI{1192}{\second} 	& \SI{373}{\second}		& \SI{372}{\second} \\
								\hline
\multirow{2}{*}{montage} 		& \SI{9}{\percent} 		& \SI{18}{\percent} 			& \SI{16}{\percent} 		& \SI{17}{\percent} 		& \SI{18}{\percent} \\
								& \SI{9}{\milli\second}	& \SI{15}{\milli\second} 	& \SI{646}{\second} 		& \SI{128}{\second}		& \SI{227}{\second} \\
								\hline
\multirow{2}{*}{soykb}		 	& \SI{1}{\percent} 		& \SI{1}{\percent} 			& \SI{2}{\percent} 		& \SI{3}{\percent} 		& \SI{3}{\percent} \\
								& \SI{4}{\milli\second}	& \SI{8}{\milli\second} 		& \SI{390}{\second} 		& \SI{16}{\second}		& \SI{15}{\second} \\
								\hline
\multirow{2}{*}{srasearch}	 	& \SI{13}{\percent} 		& \SI{16}{\percent} 			& \SI{24}{\percent} 		& \SI{22}{\percent} 		& \SI{22}{\percent} \\
								& \SI{0}{\milli\second}	& \SI{0}{\milli\second} 		& \SI{126}{\second} 		& \SI{1}{\second}		& \SI{1}{\second} \\
\end{tabular}
\label{tbl:wfcommons}
\end{table}

We evaluate the two list scheduling algorithms HEFT and PEFT, the genetic algorithm NSGA-II as well as the FirstFit variants of the two presented decomposition-based algorithms. For two of the provided benchmark sets (\code{bwa} and \code{seismology}), neither of the algorithms could find a significant acceleration compared to a pure CPU mapping. Table~\ref{tbl:wfcommons} shows the total execution times and the average relative improvement achieved by the five mapping algorithms for all benchmark sets where an acceleration was possible. Decomposition-based mapping leads to significantly higher improvements than both HEFT and PEFT. In the \code{montage} benchmark set, PEFT achieves similar improvements compared to the decomposition approaches. Here, a small number of nodes near the end of the computation are responsible for most of the makespan. The global mapping problem therefore largely reduces to mapping these tasks correctly. Having a mostly independent subset of tasks that are responsible for the majority of the required computing effort is a common occurrence in all considered benchmark sets. In consequence, the single node and the series-parallel decomposition mapping algorithms lead to similar results. Generally, this means that the latter algorithm often appears as a slightly slower variant of the former one. In the \code{epigenomics} dataset, the series-parallel approach frequently finds mappings faster than the single node approach. The workflows here primarily consist of long chains of operations, which are executed in parallel. This forms a series-parallel graph, which can be efficiently processed with a corresponding decomposition. However, the actual achieved acceleration varies strongly with the parameterization. For some graphs of this graph set, the series-parallel approach is up to four times faster than the single node approach, while for others it is up to four times slower.

The genetic algorithm generally achieves makespan improvements that are similar to those of the decomposition-based approaches, but is overall more time-intensive. Depending on the graph set, the execution time of NSGA-II is \num{3} to \num{126} times higher. However, part of this discrepancy may be explained by the fixed number of \num{500} generations, which can be assumed to be too high for simple graph sets like \code{srasearch}. At the same time, for the \code{montage} dataset, the mapping found by NSGA-II is slightly worse than the ones found by decomposition mapping approaches, indicating that the number of generations might be too low for these highly complex task graphs.




\section{Conclusion}

The task mapping problem in heterogeneous systems for complex applications with many tasks and a high number of dependencies poses a significant challenge for state-of-the-art static task mapping algorithms. Commonly used static list scheduling algorithms exhibit a too local view on the problem to be able to find non-trivial improvements in large task graphs. In contrast, mapping algorithms based on integer linear programming are too slow to be applied to large problem instances. Genetic algorithms are able to cope with complex environments, but their inherent randomness results in a high execution time compared to a more focussed approach.

Decomposition-based task mapping algorithms fill this gap by providing significant makespan improvements for complex systems in reasonable time. The proposed FirstFit heuristic has proven to be generally preferable to the basic approach as it yields similar results in much shorter time. With this heuristic, one can expect to find static task mappings at least \num{5} to \num{10} times faster compared to genetic algorithms without a significant trade-off in terms of result quality.

Using a single-node decomposition is often sufficient to find a reasonably good task mapping. However, depending on the structure of the application, using a series-parallel decomposition can be both more effective and more efficient. Our newly introduced decomposition algorithm enables the application of series-parallel decomposition mapping on generalized DAGs. Nevertheless, the highest benefit can be expected if the task graph of the application is at least almost series-parallel.

\bibliographystyle{IEEETran}
\bibliography{literature}

\end{document}